\documentstyle[prb,twocolumn,aps,epsfig]{revtex}
\begin{document}
\draft

\title{Thermodynamics of the $(1,\frac{1}{2})$ Ferrimagnet in Finite
Magnetic Fields}

\author{K. Maisinger$^{\dag}$, U. Schollw\"{o}ck$^{\dag}$, S. Brehmer$^{\ddag}$,
H.J. Mikeska$^{\ddag}$,
Shoji Yamamoto$^{\S}$}
\address{$^{\dag}$Sektion Physik, Ludwig-Maximilians-Universit\"{a}t M\"{u}nchen,
Theresienstr.\ 37, 80333 Munich, Germany \\
$^{\ddag}$Institut f\"{u}r Theoretische Physik, Universit\"{a}t Hannover,
Appelstrasse 2, 30167 Hannover, Germany \\
$^{\S}$Department of Physics, Faculty of Science, Okayama University, 
Tsushima, Okayama 700, Japan}

\date{May 28, 1998}

\maketitle
\begin{abstract}
We investigate the specific heat and magnetisation 
of a ferrimagnet with $gS=1$ and $S=\frac{1}{2}$ spins in a finite magnetic 
field using the transfer matrix DMRG down to $T=0.025J$. 
Ferromagnetic gapless and antiferromagnetic gapped excitations for $H=0$
lead to rich thermodynamics for $H \geq 0$. While the specific heat is
characterized by a generic double peak structure, magnetisation reveals
two critical fields, $H_{c1}=1.76(1)$ and $H_{c2}=3.00(1)$ with 
square-root behaviour in the $T=0$ magnetisation. Simple analytical 
arguments allow to understand these experimentally accessible findings. 
\end{abstract}
\vspace*{0.6cm}

In recent years, one-dimensional ferrimagnets with alternating spins $gS$ and 
$S$ have attracted considerable 
attention\cite{Kahn 95,Kolezhuk 97,dimer,Alcaraz 97,Yamamoto 98b,Yamamoto 98c}.
The Lieb-Mattis theorem reveals the characteristic feature of ferrimagnets:
the ground state of a ferrimagnet with $N$ elementary cells of two spins
is a macroscopic spin of length $NS(g-1)$.
While the interaction is antiferromagnetic, the ground state resembles that
of a ferromagnet with N\'{e}el-like alignment of big and small spins. Due to 
quantum fluctuations, the
classical N\'{e}el state is not exact, but
the macroscopic magnetisation of the ground state makes spin wave
theory applicable, even in one dimension.

Spin wave theory\cite{Kolezhuk 97} 
yields two types of excitations: 
starting from magnetisation $NS(g-1)$,
there are ferromagnetic (FM) gapless excitations to states with magnetisation
$NS(g-1)-1$ (Goldstone modes), and antiferromagnetic (AFM) gapped excitations
to states with magnetisation $NS(g-1)+1$. 

These two excitation types lead to a crossover
in the behaviour of the specific heat $C_v$ and the 
susceptibility $\chi$ \cite{Yamamoto 98b,Yamamoto 98c}. $C_v$ shows an 
AFM mean field peak at intermediate
temperatures and a FM $C_{v} \propto \sqrt{T}$ behaviour for
$T\rightarrow 0$.
$\chi \propto T^{-1}$ for $T\rightarrow\infty$, but shows
a FM $T^{-2}$ divergence at $T\rightarrow 0$.

At finite field all dispersion relations
are shifted and the ground state degeneracy is lifted, such that 
experimentally easily observable changes in the thermodynamic quantities 
should occur. In this paper we discuss the expected observations and provide
very precise quantitative results obtained mainly by 
the transfer matrix Density Matrix Renormalization 
Group (DMRG)\cite{White 92,Nishino 95}, 
hoping to stimulate further experimental
investigation. 

We consider the generic $(1,\frac{1}{2})$ ferrimagnet with $gS=1$ and 
$S=\frac{1}{2}$. For this case, the AFM gap in zero field is numerically
found\cite{Kolezhuk 97} to be $\Delta^{AFM}(0)=1.759 J$.

{\em Analytical results.} The introduction of a field term $H\sum_{i}S^{z}_{i}$leaves all eigenstates invariant, while shifting the eigenenergies
by $HS^{z}_{tot}$, where $S^{z}_{tot}$ is the total magnetisation of the
eigenstate. 

Linear spin wave analysis based on the classical ferrimagnetic ground state
gives the following two dispersion ``branches'' 
(all energies, temperatures and fields
are measured in units of $J\equiv 1$):
\begin{equation}
\omega^{\pm}(q) = \left( 
\sqrt{5-4\cos q} \pm 1 \right)/2 \mp H
\end{equation}
Spin wave theory gives eigenstates of $S^{z}_{tot}$; thus the only effect 
of an external field on a one-particle
excitation carrying magnetisation $\pm 1$ is the Zeeman term introducing a gap
$\Delta^{FM}(H)=H$ to the
FM excitations $\omega^{-}$, while the AFM
branch $\omega^+$ is reduced in gap: $\Delta^{AFM}(H) = \Delta^{AFM}(0) - H$.

In linear spin wave theory, $m$-particle excitations carrying magnetisation
$\pm m$ are simply linear superpositions of $m$ one-particle excitations.
As the field acts only through the Zeeman term, it is sufficient in this
approximation to
consider a one-particle excitation to study field effects. 

At $H=\Delta^{AFM}(0)$, AFM excitations 
become gapless, 
allowing
spin-up flips at no energy cost, such that the simple result $M(T=0)=0.5$
breaks down. Using the numerical value for the gap, $\Delta=1.759$, rather than
the spin wave result, we conclude that there is a first critical field $H_{c1}=
\Delta(H=0)=1.759$, above which $M(T=0)>0.5$.

Second, for a field $H=\Delta(H=0)/2$ (i.e.\ $H_{c1}/2$, low temperature 
magnetisation
$M(T)$ should be constant with $T$, 
while it is monotonically decreasing with $T$
for $H < \Delta/2$, and monotonically increasing with $T$ for
$H > \Delta/2$. This is because the dispersion relations of the harmonic
spin waves increasing and decreasing magnetisation become identical at
this field value.

Linear spin wave analysis based on the fully
polarized state yields a dispersion relation
\begin{equation}
\omega^{\pm}(q) = H -\frac{3}{2} \pm \frac{1}{2} \sqrt{5 + 4\cos q} =
H - \omega^{F\pm}_{1}(q),
\end{equation}
where $\omega^{F\pm}_{1}(q)$ is the two-branched {\em exact} 
FM {\em one-particle} excitation: both excitation branches
reduce magnetisation.
Below a second critical field $H_{c2}=3$, one dispersion branch acquires
negative energy, such that a breakdown of the
simple result $M(T=0)=3/2$ is predicted for $H<H_{c2}$.

\begin{figure}
\centering\epsfig{file=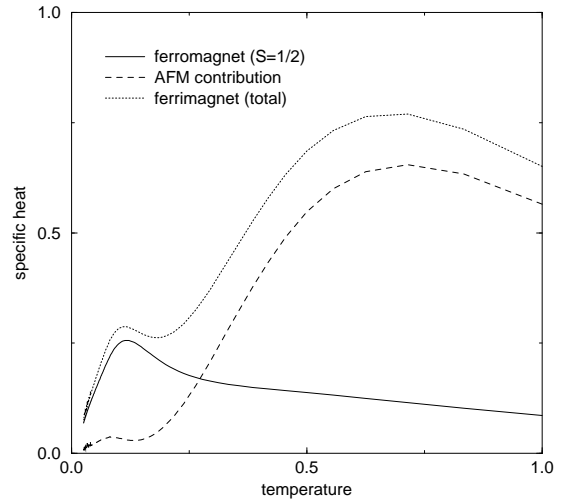,scale=0.44}
\vspace{0.3cm}
\caption{$C_{v}$ at $H=0.05$ for a ferrimagnet (dotted, DMRG) and
a $S=\frac{1}{2}$ ferromagnet (solid, DMRG). To a
very good approximation, the difference in specific heat (dashed)
can be identified as AFM contribution.}
\label{fig:decomp}
\end{figure}

Let us now consider the specific heat.
$C_{v}(T)$ should acquire a generic 
double-peak structure for low fields $0<H<H_{c1}$, 
because two gapped excitation modes exist:
Both gapped antiferromagnets as well as ferromagnets in external
fields (see Fig.\ \ref{fig:decomp}) show exponential activation of $C_v$ with a
pronounced peak, whose position is related to the gap or field
size. Peak
positions should thus shift with $H$ to higher (FM contribution)
and lower (AFM contribution) temperatures because of the
Zeeman term. At the critical fields, where the gapless excitation has the form 
$\omega \propto q^2$, $C_v \propto \sqrt{T}$ is expected for
$T\rightarrow 0$.

Another analysis for low fields can be obtained from studying the 
decoupled-dimer limit, useful in the $H=0$ 
case\cite{dimer}: every second
interaction is switched off, yielding a flat
dispersion, but the FM and AFM
elementary excitations are still gapless and gapped, respectively.
This analysis gives a double-peak structure for $C_{v}$. 
With increasing field, the
double-peak structure is smeared out and the two peaks merge into
one. With further increase of the field, 
the system changes discontinuously into the fully polarized state.

The existence of two critical fields $H_{c1}$ (up to which the ground
state magnetisation persists) and $H_{c2}$ (which marks the beginning
full polarisation) in the ferrimagnet is analogous to the finite field
behaviour of the $S=1$ (Haldane) chain. In both systems the
characteristic property of gapped excitations existing for both $0< H
< H_{c1}$ and for $H > H_{c2}$ breaks down at the critical fields; in
spin wave theory this is indicated by an instability of an arbitrary
number of spin waves (spin wave condensation). As for the $S=1$ chain
(Ref.\ \onlinecite{Affleck 90}) we expect the occurence of a critical 
(Luttinger liquid) phase in the intermediate regime $H_{c1} < H < H_{c2}$. 
This phase should resemble a (critical) anisotropic Heisenberg chain with
the anisotropy determined by the magnetic field; power law
correlation functions as well as a linear behaviour of the specific
heat $C_v = \gamma T$ (with $\gamma$ related to $H$) are expected. The
ferrimagnetic chain, however, having two spins per unit cell, is
richer than the $S=1$ chain and the interplay of two elementary
excitations leads to new crossover phenomena between high and low
field behaviour in both specific heat and magnetisation.

The spin wave approach thus results in an interesting qualitative
picture, but evidently it suffers from a number of deficiencies: The
results for the non-Zeeman part of the excitation energies from the
ferrimagnetic ground state is only approximate, the stiffness of
the FM excitations is overestimated and the gap energy of the AFM excitations
is underestimated (Ref.\ \onlinecite{Kolezhuk 97}), 
for quantitative estimates the numerical
values should be used. Spin wave theory does not yield a peak for the
specific heat as the harmonic approximation implies $dC_v/dT > 0$ for
an arbitrary density of states; however, 
for small fields, it gives an unusually high $C_v$ at very low $T$,
consistent with a second low-$T$ peak. 

The predictions of spin wave theory for the critical fields as
described above, further rest on the assumption that multi particle
excitations do not become unstable before the one particle excitation.
If this is true, as is plausible and generally asumed, the value
$H_{c2} = 3$ for the upper critical field is exact (since the
corresponding one particle excitation energy is exact); this appears
likely although a formal proof is not available.
 
{\em Transfer Matrix DMRG Results.}
To obtain quantitatively reliable values for the specific heat and the 
magnetisation, the recently proposed transfer matrix
DMRG\cite{Nishino 95} was applied to the problem. 
Essentially a decimation procedure applied
to a quantum transfer matrix, it maintains the advantages of the quantum
transfer matrix method such as working in the thermodynamical limit, and allows
the evaluation of very large Trotter numbers, giving reliable access to
thermodynamics at very low temperatures. In the problem under study, most
interesting observations can be made at very low temperature; in the critical
region $H_{c1}<H<H_{c2}$ the absence of finite size effects is of
advantage. 

The (controlled) approximation of the DMRG rests on keeping a reduced state
space. We find that keeping $M=80$ states yields almost exact results 
(note that we overestimate the specific heat by several percent at higher
temperatures, due to coarse DMRG sampling at high $T$: sampling temperatures
are spaced by $\Delta\beta=0.2$),
excluding the case $H=0$, where the very high FM degeneracy
makes it necessary to push the method much further\cite{Yamamoto 98c}.  

\begin{figure}
\centering\epsfig{file=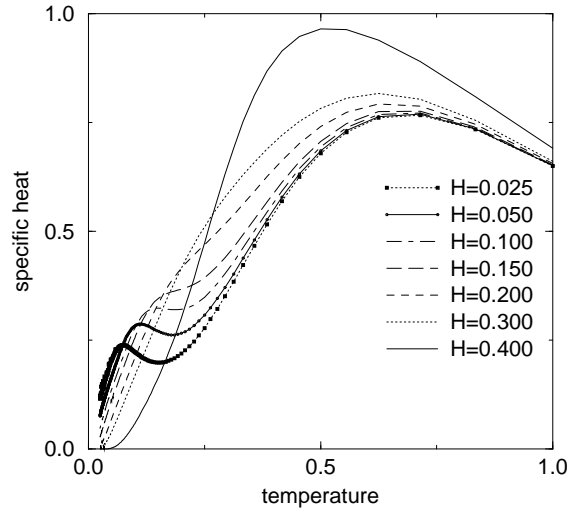,scale=0.57}
\vspace{0.3cm}
\caption{$C_{v}$ vs.\ $T$ for fields $H\leq 0.4$.} 
\label{CvoverTlowH}
\end{figure}

\begin{figure}
\centering\epsfig{file=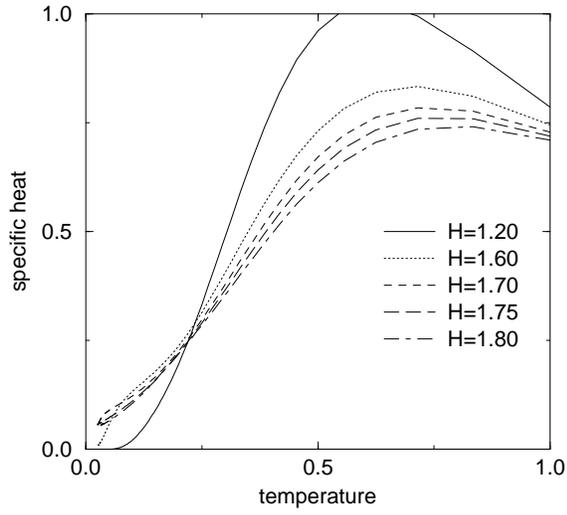,scale=0.57}
\vspace{0.3cm}
\caption{$C_{v}$ vs.\ $T$ for $H$ close to $H_{c1}=1.76(1)$.}
\label{CvoverTbigH}
\end{figure}

\begin{figure}
\centering\epsfig{file=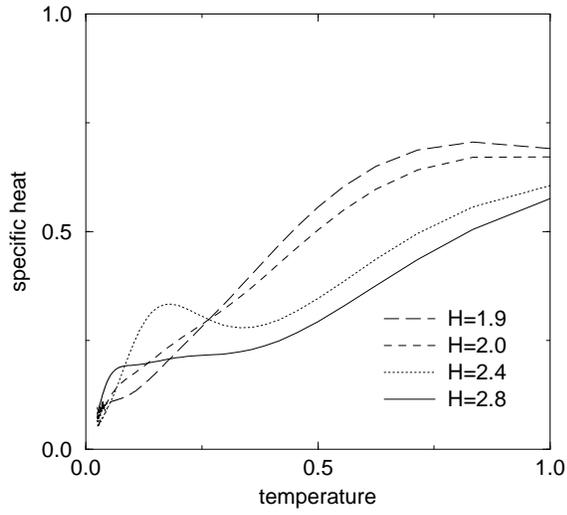,scale=0.57}
\vspace{0.3cm}
\caption{$C_{v}$ vs.\ $T$ for fields $H>H_{c1}=1.76(1)$.}
\label{CvoverThighH}
\end{figure}

In the specific heat per elementary cell, 
a two-peak structure evolves for small fields, with
the high-temperature AFM peak moving to smaller temperatures and the
low-temperature FM peak moving to higher temperatures 
(Figure \ref{CvoverTlowH}). Using for a first rough estimate
$ C_v \propto x^2 \cosh^{-2} x, \quad x = J \Delta/2T$, leading to 
$\Delta \approx 2.4 T_{peak}$, one finds for the high-$T$ peak a 
zero-field gap $\Delta_{cv}\approx 1.70 J$ 
in reasonable agreement with the
numerically observed gap ($\Delta \approx 1.759 J$); 
it shifts linearly and almost proportionally with
$H$ to lower $T$. As Fig.\ \ref{fig:decomp} shows, the low-$T$ peak is
almost identical with that of a $S=\frac{1}{2}$ ferromagnet; the ``remaining''
specific heat can be attributed to AFM excitations (this naive comparison is 
of course not perfect, but renormalization effects are surprisingly weak).
For $H > 0.2$, the two peaks merge into one; a
shoulder for $H=1.6$ indicates the emergence of a second peak (Figure 
\ref{CvoverTbigH}). For $H>H_{c1}$
a two peak structure emerges again (Figure 
\ref{CvoverThighH}). 
In the intermediate regime $H_{c1} < H < H_{c2}$ the expected linear
behaviour of the low temperature specific heat is clearly observed for
fields $H = 2.0, 2.4$. For magnetic fields close to the critical values
our results seem to indicate a behaviour $C_v \propto \sqrt{T}$ for
$T \to 0$.

\begin{figure}
\centering\epsfig{file=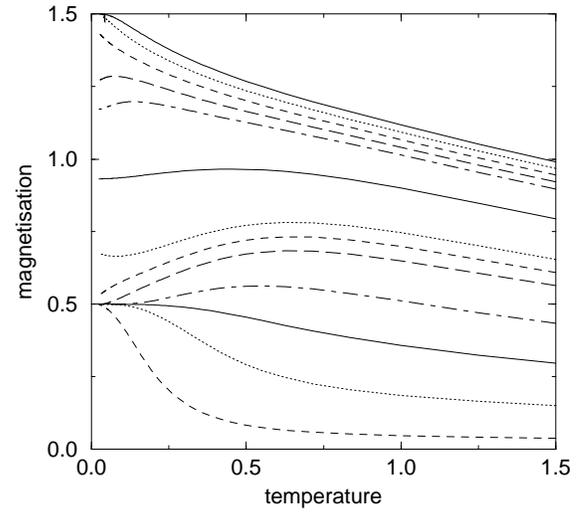,scale=0.57}
\vspace{0.3cm}
\caption{Magnetisation vs.\ $T$ for fields 
$H=0.1$, $0.4$, $0.8$, $1.2$, $1.6$, $1.75$, $1.9$, $2.4$, $2.8$, $2.9$, 
$3.0$, $3.1$, $3.2$ (from the
lowest to the highest curve).}
\label{MoverTmanyH}
\end{figure}

\begin{figure}
\centering\epsfig{file=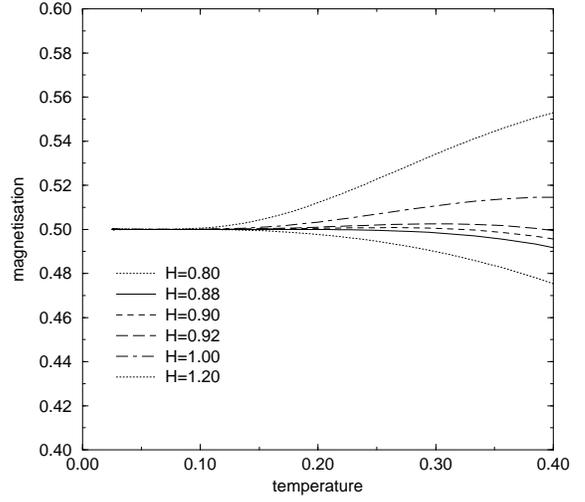,scale=0.44}
\vspace{0.3cm}
\caption{$T$-independence of the low-$T$ magnetisation 
for magnetic fields close to half the 
critical field $H_{c1}=1.76(1)$.}
\label{MvoverThalfH}
\end{figure}

Considering the magnetisation per elementary cell, 
one finds for $H<H_{c1}$ that
$M=1/2$ at $T=0$. 
How this magnetisation is reached for a fixed field $H$,
is due to a competing effect of magnetising interactions and thermal
fluctuations. For high temperature,
thermal fluctuations will always suppress magnetisation. For fields below
$H=0.88$, demagnetising FM fluctuations cost less in energy and
also suppress magnetisation. For higher fields, magnetising AFM
fluctuations will increase magnetisation. One observes therefore an intermediate
magnetisation peak before thermal fluctuations again suppress magnetisation 
(Figure \ref{MoverTmanyH}).
We observe an almost flat magnetisation curve up to $T\approx 0.2$ at $H
\approx 0.88$, in excellent accordance with the prediction 
(Figure \ref{MvoverThalfH}).  

\begin{figure}
\centering\epsfig{file=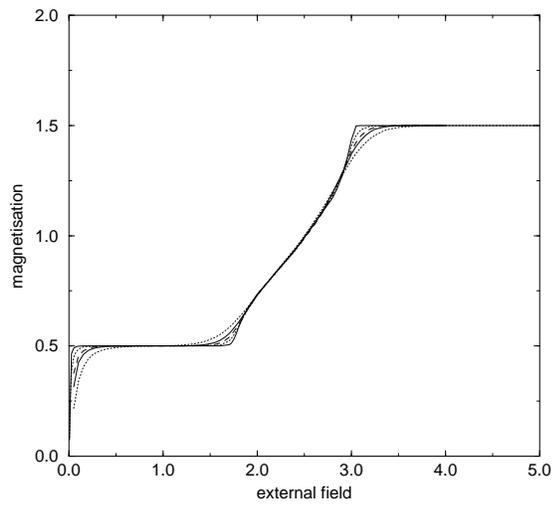,scale=0.44}
\vspace{0.3cm}
\caption{Magnetisation vs.\ $H$ for $T=0.025$, $0.05$ (DMRG), $T=0.08$, 
$0.1$, $0.15$ (quantum MC, to show consistency of
methods). Curves approach singular behaviour for $T\rightarrow 0$.}
\label{MoverHlowT}
\end{figure}

We find the lower critical field (Figure \ref{MoverHlowT})
at $H_{c1}=1.76(1)$, 
the ferrimagnetic magnetisation at $T=0$
is broken up at $H=H_{c1}$, 
with $M(H)-M(H_{c1}) \propto \sqrt{H-H_{c1}}$ (cf.\
Refs.\ \onlinecite{Affleck 90} for equivalent observations in Haldane magnets).
For $T\rightarrow 0$, we find that $(M(H)-M(H_{c1}))^2$ becomes linear in $H$
close to $H_{c1}$. As the 
system is critical at this field, working with the transfer matrix DMRG, which
is in the thermodynamic limit, is preferable to conventional DMRG, which
would suffer from finite-size effects and possible metastable 
trappings\cite{Hieida 97} 
due to extensive level crossings at $H_{c1}$ and immediately above.
Numerically, we locate the upper critical field at $H_{c2}=3.00(1)$, in
excellent agreement with the analytical result. Again, a square root
behaviour in the approach to the maximum magnetisation can be observed. 

{\em Conclusion.}
We have shown that the properties of a ferrimagnet in a weak external field can
be understood qualitatively and in some cases even quantitatively in an
extremely simple picture. The specific heat clearly reflects  
the dual structure of the excitations of the ferrimagnet for all fields. 
The magnetisation is dominated by the two critical fields, 
which can be understood
as pure Zeeman effects. The Luttinger-like critical phase in between, 
with square root
divergences in the magnetisation, needs consideration of 
many-body effects for understanding.
The presented finite-temperature
properties, in particular the low-field behavior of $C_{v}$ 
should be accessible to experimental verification.

\end{document}